\documentstyle[12pt,math]{article}
\begin{document}
\parskip=4mm
\parindent=2.5em
\renewcommand{\theequation}{\arabic{section}.\arabic{equation}}
\renewcommand{\thefootnote}{*}
\thispagestyle{empty}
\noindent October 1994 \hfill{UNIL-TP-4/94, hep-ph/9410302}

\vspace*{1cm}
\begin{center}
{\LARGE\bf Consistency of the Chiral Pion-Pion Scattering Amplitudes
with Axiomatic Constraints\footnotemark}

\vspace{8mm}

{\large B.~Ananthanarayan, D.~Toublan, G.~Wanders}\\[4mm]
Institut de physique th\'{e}orique, Universit\'{e} de Lausanne\\
CH-1015 Lausanne, Switzerland\\
{\small (e-mail: BANANTHA@ULYS.UNIL.CH)}

\footnotetext{Work supported by the Swiss National Science Foundation}

\vspace{12mm}

{\small{\bf Abstract}

\parbox{12cm}{The pion-pion scattering amplitudes provided by second-order
chiral
perturbation theory are confronted with known rigorous constraints derived
from the axioms of quantum field theory. We mainly test constraints
restricting the $\pi^0$-$\pi^0$ $S$- and $D$-wave amplitudes in the
unphysical interval $0\leq s\leq 4m_\pi^2$. These constraints impose
significant lower bounds for a linear combination of coupling constants
specifying the second order chiral Lagrangian. The accepted value of this
combination is consistent with these bounds. The $\pi^0$-$\pi^0$ $S$- and
$D$-waves are strongly correlated by a set of constraints.}}
\end{center}

\noindent{\small }

\vfill{\noindent PACS numbers: 11.10.Cd, 12.39.Fe, 25.80.Dj.

$\mbox{}$}
\newpage
\setcounter{page}{1}
\section{Introduction}

\hspace*{2.5em}During the sixties and seventies much effort was invested
in deriving properties of scattering amplitudes which are exact
consequences of the general principles of quantum field theory~\cite{ref1}.
This program was pioneered by Andr\'{e}' Martin and was very successful in the
pion-pion case, the scattering of the lightest hadrons. Analyticity
properties, i.e.~the fact that scattering amplitudes are boundary values of
analytic functions of two complex variables, constitute one of the main tools
provided by axiomatic field theory. It implies the validity of dispersion
relations with a number of subtractions restricted by the Froissart bound. The
other tools are unitarity and crossing symmetry. Their interplay leads to a
wealth of constraints on the low energy pion-pion scattering~\cite{ref2}. We
call them ``axiomatic constraints" as they follow from the axioms of quantum
field theory.

A characteristic feature of the axiomatic constraints is that they do not
depend on any specification of the interactions going beyond the requirement
of crossing symmetry. Nowadays it is well established that the pion is a
quasi Goldstone boson associated with the breaking of chiral symmetry and the
pion-pion scattering amplitudes should reflect the specificities of the
dynamics of such particles. This has been worked out in chiral perturbation
theory (CPT), which provides an approximate form of the low energy pion-pion
amplitudes. Our aim is to check whether this chiral Ansatz, i.e.~the
second-order one-loop chiral amplitudes first obtained by Gasser and
Leutwyler~\cite{ref3}, satisfies a representative set of the axiomatic
conditions. Surprisingly, this has not been done until now, at least to the
best of our knowledge.

The constraints we shall test restrict the shape of the pion-pion amplitudes
in a triangle $\Delta$ of the Mandelstam plane:
\begin{equation}
\Delta=\{s,t,u\;|\;0\leq s\leq 4m^2_\pi,\;0\leq t\leq 4m^2_\pi,\;0\leq u\leq
4m^2_\pi\},
\end{equation}
where $s$, $t$ and $u$ are the standard Mandelstam variables, $s+t+u=4m_\pi^2$,
$m_\pi=$ pion mass. As the chiral Ansatz is meant to provide a reliable
approximation of the pion-pion amplitudes at low values of $s$, $t$ and $u$,
it should verify our conditions with a good precision. In fact, the chiral
amplitudes are analytic functions, they are exactly crossing symmetric and
have positive absorptive parts. The axiomatic conditions being consequences
of these properties, they might be expected to be satisfied automatically.
This is not the case because the chiral amplitudes grow asymptotically as
$s^2$ and violate the Froissart bound which is another ingredient of the
axiomatic constraints. Since the chiral Ansatz represents the first terms of a
low
energy expansion, bad asymptotic behavior can be expected. Our purpose is to
determine whether the chiral Ansatz, when restricted to its domain of validity,
is
compatible with the low energy implications of the Froissart bound. The latter
being the mark of a local quantum field, we are asking if quasi Goldstone
bosons can be described by such a field.

Our constraints are inequalities which are linear in the amplitudes: they
enforce bounds on combinations of parameters appearing in
the chiral Lagrangian. This implies that these quantities cannot
be chosen at will if compatibility with general field theoretic principles is
required. In order to get a first insight into the nature of these
restrictions, we apply the constraints to the standard chiral perturbation
theory defined by Gasser and Leutwyler~\cite{ref3}. This may as well be done
for other
versions of this theory, for instance the ``generalized chiral perturbation
theory" proposed by Stern and his collaborators~\cite{ref}. In the case of the
standard theory, we find that a combination $\bar{l}$ of second order coupling
constants
defined in~(2.7) has to be larger than a lower bound equal to
6 in order of magnitude. Fits to
experimental data give $\bar{l}\approx 21$~\cite{ref4}. The bound is respected
and the order of magnitude of the bound is not disproportionate to the actual
value of $\bar{l}$. This is of importance because it proves that the
restrictions imposed by the constraints are relevant, a fact which could not
have been
asserted beforehand.  A sum rule requiring a phenomenological input accounts
for
the difference between the experimental value of $\bar{l}$ and its axiomatic
lower bound.

Since the axiomatic constraints are a rather old topic, we find it worthwhile
to start with an outline of their sources. This is done in Section~2. The
standard chiral Ansatz for the $\pi^0$-$\pi^0$ amplitude is also displayed in
this Section and first co
conditions on $S$ and $P$ waves and bounds for $\bar{l}$ are obtained, taking
into account the uncertainties due to unknown third-order corrections.
The sum rule which has just been mentioned is also discussed in this Section.
Section~4 is devoted to constraints which strongly correlate the
$\pi^0$-$\pi^0$ $S$- and $D$-waves. Our conclusions are presented in Section~5.

\setcounter{equation}{0}
\section{Axiomatic properties of pion-pion scattering}

\hspace*{2.5em}We first recall the basis of the rigorous properties of the
pion-pion amplitudes we shall exploit~\cite{ref1}. These properties hold in
the triangle $\Delta$ defined in (1.1) where the amplitudes are real and
satisfy the crossing conditions
\begin{eqnarray}
T^I(s,t,u)&=&\sum_{I'}C_{st}^{II'}\,T^{I'}(t,s,u)\;=\;
\sum_{I'}C_{su}^{II'}\,T^{I'}(u,t,s) \nonumber\\
&=&\sum_{I'}C_{tu}^{II'}\,T^{I'}(s,u,t).
\end{eqnarray}
$T^I(s,t,u)$ is the $s$-channel isospin $I$ amplitude ($I=0,1,2$) and
$C_{st}$, $C_{su}$, $C_{tu}$ are crossing matrices~\cite{ref2}. At each point
of $\Delta$, $T^I(s,t,u)$ is given by three dispersion relations evaluated
either at fixed $s$, fixed $t$ or fixed $u$. These relations result from exact
analyticity properties: the Froissart bound ensures that only two subtractions
are needed. The fixed-$t$ dispersion relations have their simplest form if
written for the isospin $I$ $t$-channel amplitudes $T^{(I)}(s,t,u)$
($=T^I(t,s,u)$):
\begin{eqnarray}
\hspace*{-2.5em}T^{(I)}(s,t,u)&=&\mu_I(t)+\nu_I(t)(s-u)\nonumber\\
&&\quad+{1\over \pi}\int_{4m_\pi^2}^\infty{\rm
d}x{1\over x^2}\left({s^2\over x-s}+(-1)^I{u^2\over x-
u}\right)\sum_{I'}C_{st}^{II'}\,A^{I'}(x,t),
\end{eqnarray}
where $\mu_I(t)$ and $\nu_I(t)$ are unknown $t$-dependent subtraction
constants ($\mu_1=0$, $\nu_0=\nu_2=0$). $A^I$ is the absorptive part of the
$s$-channel amplitude $T^I$:
\begin{equation}
A^I(s,t)=\Im T^I(s+{\rm i}\epsilon,t,u),\qquad s\geq 4m_\pi^2.
\end{equation}
If $(s,t,u)\in\Delta$, $T(s,t,u)$ is also given by a fixed-$s$ or a fixed-$t$
dispersion relation obtained from (2.2) by suitable substitutions. An
important property of the absorptive parts is that they are positive for
$0\leq t<4m_\pi^2$. This follows from their partial wave expansion,
\begin{equation}
 A^I(s,t)=\sum_{l=0}^\infty(2l+1)\,\Im f_l^I(s+{\rm
i}\epsilon)\,P_l\left(1+{2t\over s-4m_\pi^2}\right).
\end{equation}
Each term in the right hand side is positive because the partial waves
$f_l^I(s)$ have positive imaginary parts in the physical domain $s\geq
4m_\pi^2$,
and $P_l(z)\geq 1$ for $z\geq 1$. Our constraints are consequences of the
twice subtracted dispersion relations, the positivity of the absorptive parts
and crossing symmetry. They restrict the shape of the amplitudes $T^I$ in the
triangle $\Delta$ and the partial waves $f_l^I$ in the unphysical interval
$0\leq s\leq 4m^2_\pi$.

Most of the constraints we shall use concern the totally symmetric
$\pi^0$-$\pi^0$ amplitude $T=(1/3)(T^0+2T^2)$. The second-order chiral Ansatz
of this amplitude is~\cite{ref3,ref5}
\begin{eqnarray}
T_\chi(s,t,u)&=& {\lambda\over 2}+{\lambda^2\over 4\pi}\left\{\left[(2s^2-
4s+3)I(s)+(s\to t)+(s\to u)\right]\right.\nonumber\\[2mm]
&&\qquad\qquad +\left.{1\over 3}\left[\alpha(s^2+t^2+u^2)+\beta\right]\right\},
\end{eqnarray}
with $\lambda=m_\pi^2/(16\pi F_{\pi}^2)$,
where $F_\pi$ is the pion decay constant. We treat $\lambda$ as an expansion
parameter : $\lambda=0.0448$ if one takes $m_\pi=140$~MeV and
$F_{\pi}=93.2$~MeV.
In (2.5) and in the following the variables $s,\ t,\ u$ are
measured in units of $m_\pi^2$.
 Our normalization of scattering
amplitudes differs from the one commonly used in CPT by a factor ($1/32\pi$).
The constant first-order term is the $\pi^0$-$\pi^0$ Weinberg amplitude. The
first square bracket gives the finite part of the one-loop contributions with
\begin{equation}
I(s)=2\left[\sqrt{(4/s)-1}\left(\mbox{Arctan}\,\sqrt{(4/s)-1}-{\pi\over
2}\right)+1\right].
\end{equation}
This form is adapted to the unphysical values $0\leq s\leq 4$. The second
square bracket in~(2.5) comes from the tadpoles and second-order trees. The
constants $\alpha$ and $\beta$ are linear combinations of the coupling
constants $\bar{l}_i$ of the second-order chiral Lagrangian
\begin{equation}
\begin{array}{ll}
\alpha=\bar{l}-6,&\bar{l}=2\bar{l}_1+4\bar{l}_2,\\[4mm]
\beta=\bar{l}\,'+21, &\bar{l}\,'=-8\bar{l}_1-16\bar{l}_2-
9\bar{l}_3+12\bar{l}_4.
\end{array}\end{equation}
The exact $\pi^0$-$\pi^0$ amplitude $T(s,t,u)$ satisfies a simple version of
the dispersion relation (2.2):
\begin{equation}
T(s,t,u)=\mu(t)+{1\over \pi}\int_4^\infty{\rm d}x\,{1\over x^2}\left({s^2\over
x-s}+{u^2\over x-u}\right)A(x,t),
\end{equation}
where $A(s,t)$ is the $s$-channel $\pi^0$-$\pi^0$ absorptive part: it is
positive for $s\geq 4$, $0\leq t<4$.

With one exception, at the end of Section~3, the constraints we shall discuss
concern $\pi^0$-$\pi^0$ scattering and we want to know the conditions under
which they are obeyed by the chiral amplitude $T_\chi$. More precisely, we
assume that the true amplitude $T$ differs from $T_\chi$ by $O(\lambda^3)$
terms in the triangle $\Delta$:
\begin{equation}
T(s,t,u)=T_\chi(s,t,u)+O(\lambda^3),\qquad (s,t,u)\in\Delta.
\end{equation}
If the amplitude $T$ satisfies an axiomatic condition it can be violated by
$T_\chi$ to an order of magnitude fixed by equation~(2.9). Clearly, we have no
control of the third order corrections within the present context. The actual
size
of the $O(\lambda^3)$ term in (2.9) could well be $\lambda^3$ times a
relatively
large factor.

The axiomatic constraints we shall use follow from analyticity, positivity of
absorptive parts and crossing. These conditions being linear-convex and
homogeneous, they constrain only the shape of the amplitudes and not their
size. However, by including nonlinear aspects of unitarity, it has been
possible to derive absolute axiomatic bounds for the $\pi^0$-$\pi^0$
amplitude~\cite{ref6}. These bounds are remarkable in that they have to hold
independently of the details of the dynamics, whenever the $I=0$ and $I=2$
mass spectra start at $2m_\pi$ (absence of two pions bound states). The most
stringent bounds are
\begin{equation}
T(3,2,-1)>-1.30,\qquad T(4/3,4/3,4/3)<2.70.
\end{equation}
With $\lambda=0.0448$ and reasonable values of $\alpha$ and $\beta$, $T_\chi$
is of the order of a percent of the bounds. This means that the chiral Ansatz
describes a pion-pion interaction which is weak at the scale defined by the
axiomatic bounds~(2.10). It is precisely one of the achievements of current
algebra and CPT to explain the smallness of the pion-pion interaction.

{}From now on we consider only constraints derived from the linear and
homogeneous conditions of analyticity, positivity and crossing. They leave the
first-order Weinberg amplitudes completely free. This comes from the fact that
these amplitudes are linear in $s$, $t$ and $u$ and that the arbitrariness in
the subtraction constants in (2.2) leaves crossing symmetric linear terms of
the full amplitudes $T^I$ undetermined. Consequently, all conditions we shall
examine test exclusively the second-order chiral Ansatz.

As already mentioned in the introduction, the constraints lead to bounds for
linear combinations of the second-order coupling constants $\bar{l}_i$, the
bounds being determined by the loop and tadpole terms. To establish the
relevance of the these bounds we look at the simplest axiomatic constraint on
the $\pi^0$-$\pi^0$ amplitude $T(s,t,u)$. It tells us that the symmetry point
$s=t=u=4/3$ is an absolute minimum of $T$~\cite{ref7}. In fact $T_\chi$ has
an extremum at this point because of crossing symmetry. Inspection of (2.5)
shows that it will be a maximum, instead of a minimum, if $\alpha$ is large
and negative. Therefore $\alpha$ has to be larger than some lower bound. At
the symmetry point,
\begin{equation}
{\partial^2T_\chi\over \partial s^2}={\partial^2T_\chi\over \partial
t^2}={1\over 2}\,{\partial^2T_\chi\over \partial s\partial t}={\lambda^2\over
4\pi}\left[{2\over 3}\alpha+1.729\right].
\end{equation}
In so far as the second-order derivatives of the $O(\lambda^3)$ term in (2.9)
are also $O(\lambda^3)$, the symmetry point will be a minimum of $T$ if
\begin{equation}
\bar{l}>3.4+6\pi O(\lambda)=3.4+O(0.8).
\end{equation}
This condition shows that $\bar{l}_i=0$ is excluded: second-order trees have
to be included in the chiral Lagrangian in order to get an Ansatz which is
compatible with the axiomatic constraints.

The actual value of $\bar{l}$ quoted in the introduction, $\bar{l}=21\pm 4$,
comes from $\bar{l}_1=-1.7\pm1.0$ and $\bar{l}_2=6.1\pm0.5$~\cite{ref4}. It is
compatible with (2.12) and, as announced, the order of magnitude of the bound
is comparable with this accepted value.

\setcounter{equation}{0}
\section{$S$- and $P$-waves, mainly $\pi^0$-$\pi^0$ $S$-wave}

\hspace*{2.5em}We investigate two sets of axiomatic constraints on the
$\pi^0$-$\pi^0$ $S$-wave $f_0(s)$. The first set restricts the shape of $f_0$
on the interval $[0,4]$ through the signs of its derivatives. The second set
consists of inequalities relating the values of $f_0$ at two points of
$[0,4]$. For $0\leq s\leq 4$, $f_0(s)$ is given by
\begin{equation}
f_0(s)={2\over 4-s}\int_0^{(4-s)/2}{\rm d}t\,T(s,t,u).
\end{equation}

The constraints on the derivatives follow directly from properties of
$T(s,t,u)$ implied by the dispersion relation~(2.8) and the positivity of
$A(x,t)$. One finds~\cite{ref7,ref8,ref9}
\begin{eqnarray}
{{\rm d}f_0(s)\over {\rm d}s}<0 &\mbox{ for }& 0<s<1.217\\
{{\rm d}f_0(s)\over {\rm d}s}>0 &\mbox{ for }& 1.697<s<4\\
{{\rm d}^2f_0(s)\over {\rm d}s^2}>0 &\mbox{ for }& 0<s<1.7.
\end{eqnarray}
These conditions show that $f_0$ has a minimum in the interval
$(1.217,1.697)$: this minimum is clearly a reflexion of the minimum of the
full amplitude at the symmetry point.

The chiral Ansatz for the $\pi^0$-$\pi^0$ $S$-wave is
\begin{eqnarray}
\hspace*{-2.5em}f_0^\chi(s)&=&{\lambda\over 2}+{\lambda^2\over
4\pi}\left\{\left[(2s^2-
4s+3)I(s)+{2\over 4-s}\int_0^{4-s}{\rm d}t(2t^2-4t+3)I(t)\right]\right.
\nonumber\\
&&\qquad +\left.{1\over 3}\left[{\alpha\over 3}
(5s^2-16s+32)+\beta\right]\right\}.
\end{eqnarray}

We ask if this Ansatz satisfies the conditions (3.2-4) as it stands, ignoring
the $O(\lambda^3)$ corrections in (2.9). As in the case of the full chiral
amplitude,
$f_0^\chi$ will have a maximum instead of a minimum if $\alpha$ is too
negative. The minimum value of $\alpha$ is determined by the shape of the loop
contribution, given by the first square bracket in (3.5). This contribution is
found to satisfy conditions (3.2) and (3.5) but it marginally violates
condition (3.3) because its minimum is slightly above $s=1.697$, at $s=1.701$.
The polynomial in the second square bracket, coming from the tadpoles and
second-order trees, has its own extremum at $s=1.6$. Thus $\alpha$ has to be
slightly positive in order to bring the minimum of $f_0^\chi$ into the allowed
interval: this corresponds to
\begin{equation}
\bar{l}>6.63.
\end{equation}
The situation is illustrated in Fig.~3.1 which shows the evolution of the
shape of $f_0^\chi$ as $\bar{l}$ varies between 0 and 8. Fig.~3.2 shows
$f_0^\chi$ for the central phenomenological value $\bar{l}=21$.

Our findings about the loop term are instructive because the loop contribution
to the full amplitude verifies all the required exact properties except the
Froissart bound and its $S$-wave projection violates the conditions (3.2-4)
only weakly. This shows that the axiomatic constraints are not necessarily
very sensitive to wrong asymptotic behavior. In the present case, the
practical effect of these constraints is to impose correct behavior of the
tree and tadpole contributions.  Note that the third-order corrections
introduce an uncertainty $O(1.7)$ into (3.6).

The constraints relating the values of $f_0(s)$ at two points in $[0,4]$ are
obtained by eliminating the subtraction constant $\mu(t)$ in (2.8). Projecting
this dispersion relation onto the $t$-channel $S$-wave $f_0(t)$ gives an
equation relating $\mu(t)$ and $f_0(t)$. Using this equation, $\mu(t)$ can be
eliminated from (2.8) in favor of $f_0(t)$, giving
\begin{equation}
T(s,t,u)=f_0(t)+{1\over \pi}\int_4^\infty{\rm d}x\,A(x,t)\,F(x,s,t),
\end{equation}
with
\begin{equation}
F(x,s,t) ={1\over x-s}+{1\over x-4+s+t}+{2\over 4-t}\ln\left({x-4+t\over
x}\right).
\end{equation}
As $T(s,t,u)=T(t,s,u)$, (3.7) implies that
\begin{equation}
f_0(s)-f_0(t)={1\over \pi}\int_4^\infty{\rm d}x\left[A(x,t)\,F(x,s,t)-
A(x,s)\,F(x,t,s)\right].
\end{equation}

We see that the inequality
\begin{equation}
f_0(s)-f_0(t)>0
\end{equation}
holds for every pair $(s,t)$ such that
\begin{eqnarray}
\hspace*{-2.5em}F(x,s,t)>0 \mbox{ and } (F(x,s,t)-F(x,t,s))>0 &\mbox{for}&x\geq
4 \mbox{ if
}t>s,\nonumber\\[-2mm]
&&\\[-2mm]
\hspace*{-2.5em}F(x,t,s)<0 \mbox{ and } (F(x,s,t)-F(x,t,s))>0 &\mbox{for}&x\geq
4 \mbox{ if
}s>t.\nonumber
\end{eqnarray}

The inequality $A(x,s)>A(x,t)$, valid if $4>s>t\geq 0$, has been taken into
account: it results from (2.4). The first known inequalities
(3.10)~\cite{ref8} have been obtained from the more restrictive condition
\begin{equation}
F(x,s,t)>0 \quad\mbox{ and }\quad F(x,t,s)<0 \quad\mbox{ for } x\geq 4.
\end{equation}
We have computed anew the domain of the $(s,t)$-plane defined by the
condition  (3.12) and determined the significantly larger domain defined by
condition (3.11). The result is displayed in Fig.~3.3.

The assumption (2.9) implies that the inequality (3.10) imposes the following
constraint on the second order chiral $S$-wave:
\begin{equation}
f_0^\chi(s)-f_0^\chi(t)>O(\lambda^3).
\end{equation}

Using the expression (3.5) of $f_0^\chi$, we see that (3.13) is equivalent to
an inequality of the form
\begin{equation}
a(s,t)\bar{l}-b(s,t)>O(\lambda),
\end{equation}
where $a(s,t)>0$ and we consider those pairs for
which $b(s,t)>0$. These give
lower bounds for $\bar{l}$. It turns out that the loop term in (3.5) satisfies
all the inequalities arising from the strong condition~(3.12). This means that
the lower bounds produced by these inequalities are smaller than 6. The pair
$s=2.30$, $t=1.08$, is an example verifying (3.11) and not (3.12) and leading
to an equality (3.10) slightly violated by the loop term. With $a=0.0097$ and
$b=0.0588$, this gives
\begin{equation}
\bar{l}>6.1+O(5).
\end{equation}
The large uncertainty in this lower bound comes from the fact that, for
reasonable values of $\bar{l}$, the difference
$\left(f_0^\chi(s)-f_0^\chi(t)\right)$
is small compared with $\lambda^2$ and is, in fact,
$O(\lambda^3)$. Therefore $a$ is small and the ratio $\lambda/a$ is large. All
inequalities leading to interesting ratios $b/a$ follow this trend.
Furthermore
the precise numbers appearing in bounds like (3.15) depend on the
fine details of the chiral Ansatz. There are
pairs which lead to slightly larger bounds but $s$ and $t$ are very close in
these pairs and it would be unrealistic to assume that the correction to
$\left(f_0^\chi(s)-f_0^\chi(t)\right)$ is effectively of order $\lambda^3$.
Assuming that the derivative of the correction to $f_0^\chi$ is also
$O(\lambda^3)$, we may safely suppose that the correction to the difference is
$O((s-t)\lambda^3)$. If we adopt this procedure, the pair $s=1.720$, $t=1.675$
with $a=0.000388$ and $b=0.00255$ gives the largest lower bound combined with
a relatively small uncertainty:
\begin{equation}
\bar{l}>6.6+O(5).
\end{equation}

Although the bounds (3.15) and (3.16) are not disportionate to the
phenomenological value of $\bar{l}$, the difference between this
value and the bound nevertheless is relatively large and has to be
explained.  Comparing the equality (3.9) and the inequality
(3.10) we see that this difference is determined by the integral
in the right hand side of (3.9).  This integral cannot be
evaluated with the help of the chiral Ansatz alone and the explanation
we are looking for has to be based in part on the phenomenological
analysis of pion-pion scattering.  In other words we are no
longer confronting the chiral Ansatz with rigorous constraints
alone but are using (3.9) as a sum rule.  To estimate the right
hand side of (3.9) we cutoff the integral at an energy of
$1400 \ {\rm MeV} \ (x=100)$ and replace the absorptive parts by
their $S$-wave contributions leading to the integral:
\begin{eqnarray}
& D(s,t)={1 \over \pi}\int_4^{100}dx \ [F(x,s,t)-F(x,t,s)]&\nonumber\\
& {1 \over 3}\sqrt{{x \over x-4}}[\sin^2\delta^0_0(x)+2\sin^2\delta^2_0(x)]&
\end{eqnarray}

We expect $D(s,t)$ to be a good approximation of the right hand side integral
of
(3.9).  In fact, if $(s,t)$ verifies (3.11), $D(s,t)$ is smaller
than this integral and we have an improved version of (3.14):

\begin{equation}
a(s,t)\bar{l}-b(s,t)>{1\over \lambda^2} D(s,t) +O(\lambda)
\end{equation}

To evaluate $D(s,t)$ we take the phase shifts obtained from
the chiral Ansatz between threshold and $600\ {\rm MeV}$
according to the procedure described in \cite{ref5}, for the
central values of the coupling constants~\cite{ref4,ref5}.  Above $600\ {\rm
MeV}$
and up to $1400\ {\rm MeV}$ where the chiral amplitudes can no
longer be trusted and we take the $I=0$ phase shift adopted
by the Particle Data Group\cite{pdg} and the $I=2$ phase
shift given by B. R. Martin {\it et al.}\cite{mms}.  We
find $D(2.30,1.08)=2.31\times 10^{-4}$ and $D(1.720,1.675)
=8.69\times 10^{-6}$.  These results lead to the following
lower bounds which, according to (3.18) replace (3.15)
and (3.16):
\begin{eqnarray}
& \bar{l}>17.9+O(5) & \nonumber \\
& \bar{l}>17.7+O(5) &
\end{eqnarray}

The coincidence of these two bounds obtained from quite different
$(s,t)$-pairs is striking.  It supports the assumption that
$D(s,t)$ is a good approximation of the integral in (3.9).
This allows us to turn the inequalities (3.19) into approximate
equalities, that is into approximate versions of the sum rules
(3.9).  Dropping the third order corrections we notice that the
second order chiral Ansatz is not quite consistent with the sum
rules:  the right hand side has been evaluated assuming
$\bar{l}=21$.  We shall not enter into a detailed analysis of
this issue taking into account the uncertainties
on the $\bar{l}_i$.  We limit
ourselves only to the conclusion that matters here:  the
size of $D(s,t)$ which forces $\bar{l}$ to be substantially
larger than the rigorous lower bounds (3.15) and (3.16).

A refinement of the technique leading to the $\pi^0$-$\pi^0$ constraints
(3.10) produces inequalities involving the $I=0$, 1 and 2 $S$- and $P$-
waves~\cite{ref10}. We discuss only one of these constraints:
\begin{equation}\begin{array}{l}
f_0^0(s)-f_0^2(s)-1.693269f_1^1(s)\\[4mm]
\qquad-{1\over 3}\left(f_0^0(t)+2f_0^2(t)\right)+4.751676f_1^1(t)>0,
\end{array}
\end{equation}
with $s=2.6097$, $t=1.0873$. A high accuracy in the coefficients of $f_1^1$ is
mandatory in order to ensure a vanishing contribution of the first-order linear
amplitudes. After insertion of the chiral Ansatz of the $S$- and $P$-waves,
the left-hand side of (3.20) becomes a linear function of $\bar{l}_1$,
$\bar{l}_2$ and $\bar{l}_4$. Remarkably, it depends practically only on the
combination $\bar{l}=2\bar{l}_1+4\bar{l}_2$ which appears in the
$\pi^0$-$\pi^0$ amplitudes. Eliminating $\bar{l}_2$, (3.20) becomes
\begin{equation}
0.0338\bar{l}+1.3\cdot 10^{-3}\bar{l}_1 -6\cdot 10^{-7}\bar{l}_4-
0.0764>O(\lambda).
\end{equation}
This inequality effectively constrains only $\bar{l}$:
\begin{equation}
\bar{l}>2.3+O(1.3).
\end{equation}
This bound is weaker than the preceding ones. It may be that the uncertainty
is underestimated because it is a combination of the $\lambda^3$-corrections
of the six amplitudes appearing in the constraint.

\setcounter{equation}{0}
\section{$\pi^0$-$\pi^0$ $S$- and $D$-wave}

\hspace*{2.5em}The partial waves $l\geq2$ do not have the same status as the
$S$- and $P$-waves. The reason is that the twice-subtracted fixed-$s$
dispersion relations lead to a Froissart-Gribov representation for these
higher partial waves, which involves only absorptive parts and no
subtraction constants. For the $\pi^0$-$\pi^0$ $D$-wave and $0\leq s<4$,
\begin{equation}
f_2(s)={4\over 4-s}{1\over \pi}\int_4^\infty{\rm
d}x\,A(x,s)\,Q_2\left({2x\over 4-s}-1\right).
\end{equation}
On the other hand, the chiral Ansatz gives
\begin{eqnarray}
f_2^\chi(s)&=&{\lambda^2\over 4\pi}\left[{2\over 4-s}\int_0^{4-s}{\rm d}t(2t^2-
4t+3)P_2\left(1-{2t\over 4-s}\right)I(t)\right.\nonumber\\
&&\qquad\left.+{1\over 45}\alpha(4-s)^2\right].
\end{eqnarray}
Positivity immediately implies that $f_2(s)$ as given by (4.1) is positive on
$[0,4]$. Moreover, it has been shown that
${\rm d}f_2(s)/{\rm d}s$ is negative for $1.435\leq s<4$~\cite{ref11}.

The loop contribution to the chiral $D$-wave is marginally in conflict with
these constraints, being slightly negative above $s=2.71$. It is found that
$f_2^\chi$ has the correct shape if $\bar{l}>7.85$. Although this is our
largest rigorous lower bound, it is not really useful because there is no
reliable way
of estimating the uncertainty coming from the third-order $D$-wave
corrections. The shape of $f_0^\chi$ for various values of $\bar{l}$ is shown
in Fig.~4.1.

Apart from the properties of the $D$-wave alone, there are two known sets of
inequalities involving both the $\pi^0$-$\pi^0$ $S$- and
$D$-waves~\cite{ref12}. The inequalities of the first set are of the form
\begin{equation}
C(s,t)f_2(s)+C(t,s)f_2(t)\mbox{ \raisebox{-1ex}{$\d\stackrel{\d >}{<}$} }
f_0(s)-f_0(t)
\end{equation}
for appropriate pairs $(s,t)$, $C(s,t)$ being a known function. These
inequalities again provide bounds for $\bar{l}$ but they do not improve our
previous results.

The second set of constraints gives upper or lower bounds for the $D$-wave at
given points in terms of the difference of the $S$-wave between two
points:
\begin{equation}
f_2(s)\mbox{ \raisebox{-1ex}{$\d\stackrel{\d >}{<}$} }
\left[5P_2\left(1-{2t\over 4-s}\right)\right]^{-
1}\left(f_0(s)-f_0(t)\right).
\end{equation}

These inequalities impose lower bounds on $\bar{l}$ as well as
upper bounds.  The lower bounds are weaker than the previous
ones and the upper bounds are too large ($O(1000)$) to be
of any interest.  The real
strength of the inequalities (4.4) is that they strongly constrain
the shape of the $\pi^0$-$\pi^0$ $D$-wave once the $S$-wave is given. This
phenomenon has already been recognized by Martin in a general
context~\cite{ref12}. In our case, with $\bar{l}=21$, we obtain an impressive
picture,
displayed in Fig.~4.2.
The chiral $S$-wave defines gates through which the $D$-wave must pass. The
gates located below $s=1$ are very narrow. The chiral $D$-wave passes through
all the gates, often close to the lower edge. Therefore the second-order
chiral $\pi^0$-$\pi^0$ $S$- and $D$-waves obey all the constraints~(4.4). This
picture remains qualitatively unchanged if the value of $\bar{l}$ is reduced.
It is only for $\bar{l}=3.7$ that one of the bounds starts to be
violated. The constraints (4.4) also impose correlations on the third-order
corrections, due to the fact that the width of the narrow gates is a fraction
of $\lambda^3$, as shown in Table~4.1. Thus, one may expect third-order
$S$-wave corrections to be larger than the widths of the narrow gates: they
will be shifted by these corrections and the $D$-wave must comply with these
shifts.

\section{Conclusions}

We have checked whether the second-order chiral pion-pion amplitudes obey a
set of known axiomatic constraints. We have found that this is the case as
long as the value of the combination $\bar{l}$ of second-order coupling
constants is larger than about 6. As $\bar{l}\approx 21$, the constraints are
actually satisfied. It is remarkable that the best bounds~(3.6) and (3.16)
produced by different families of $\pi^0$-$\pi^0$ constraints are nearly
equal, and slightly larger than 6. This arises from the fact that the loop
terms either obey the constraints or violate them marginally. In the latter
case, a small positive $\alpha$ removes the violation. Since
$\alpha=\bar{l}-6$ is the coefficient of polynomials in (2.5) and (3.5) coming
from tadpoles and second-order trees, $\alpha\approx 0$ essentially means that
the trees nearly cancel the tadpoles.  A sum rule involving data at energies
which are above the domain of validity of the chiral Ansatz shows that
$\bar{l}$ has to be substantially larger than its axiomatic lower bound.
It is remarkable that the
constraint~(3.20), involving all $S$- and $P$-waves, effectively restricts
only that combination $\bar{l}$ which appears in the $\pi^0$-$\pi^0$
amplitudes.

Finally, we have discovered that the second-order chiral $\pi^0$-$\pi^0$ $S$-
wave practically fixes the $D$-wave below $s=1$ by means of a set of axiomatic
constraints. Surprisingly, the second-order chiral $D$-wave agrees completely
with all the conditions imposed by the $S$-wave if $\bar{l}$ is large enough.
This implies strong correlations between the third-order $S$- and $D$-wave
corrections.

\section{Acknowledgements}

We acknowledge valuable conversations with J. Gasser and
H. Leutwyler.  We thank Z. Domanski for computational help.

\newpage
\section*{Table Caption}
\begin{enumerate}
\item[Table 4.1] The chiral $\pi^0$-$\pi^0$ $D$-wave $f_2^\chi/\lambda^3$
and its
lower and upper bounds from (4.4) for $\bar{l}=21$.
\end{enumerate}

\bigskip

\section*{Figure Captions}
\begin{enumerate}
\item[Fig. 3.1] The chiral $\pi^0$-$\pi^0$ $S$-wave as given in (3.5) with
$\bar{l}\,'=0$ ($\beta=21)$, for $s\in[0,4]$ and various values of
$\bar{l}=\alpha+6$ with the two vertical lines delimiting the
position of the axiomatic minimum following from (3.2-4).
\item[Fig. 3.2] The chiral $\pi^0$-$\pi^0$ $S$-wave on the interval $[0,4]$ for
central
values of the parameters $\bar{l}$ and $\bar{l}\,'$ defined in (2.7):
$\bar{l}=21$~\cite{ref4}, $\bar{l}\,'= -58.5$~\cite{ref4,ref5}.
\item[Fig. 3.3] The domains of pairs $(s,t)$ for which an inequality (3.10)
holds true.
The black domain contains the pairs satisfying the condition (3.12). The domain
in grey are the extensions obtained by replacing (3.12) by (3.11).
\item[Fig. 4.1] The chiral $\pi^0$-$\pi^0$ $D$-wave on the interval $[0,4]$ for
various
values of $\bar{l}$.
\item[Fig. 4.2] The upper and lower bounds imposed on the $\pi^0$-$\pi^0$
$D$-wave by
the chiral $\pi^0$-$\pi^0$ $S$-wave according to the inequalities (4.4) for
$\bar{l}=21$. The orientation of the arrows distinguishes the upper from the
lower bounds. The chiral $D$-wave satisfies all constraints.
\end{enumerate}

\newpage
\begin{center}
\begin{small}
\begin{tabular}{|c|c|c|c|}
\hline
s & (lower bounds)/$\lambda^3$ & $f_2^{\chi}/\lambda^3$ & (upper
bounds)/$\lambda^3$\\
 \hline
0.0341 && 15.25609 & 15.82205\\
0.073 & 14.45934 & 14.68568 &\\
0.304 && 11.99660 & 12.12449\\
0.325 & 11.74986 & 11.78877 &\\
0.572 & 9.47923 & 9.63069 & 10.66181\\
0.589 & 5.58449 & 9.49839 & \\
0.747 && 8.34895 & 8.36308\\
0.803 && 7.97307 & 8.33016 \\
0.826 & 7.81548 & 7.82304 & \\
1.000 & 6.58737 & 6.76339 & 9.11861 \\
1.200 & 5.57584 & 5.69112 & 9.9305 \\
1.400 & 4.54194 & 4.7524 & 10.4335 \\
1.435 & 3.53843 & 4.60094 & 10.4899 \\
1.500 & 3.96047 & 4.32754 & 10.5696 \\
1.600 & 3.43562 & 3.92992 & 10.6285 \\
1.800 & 0.930242 & 3.21038 & 10.5136 \\
1.900 & 0.286525 & 2.88582 & 5.05084 \\
1.950 & 0.475615 & 2.73184 & 4.30503 \\
2.000 & 0.708817 & 2.5832 & 3.88746 \\
2.050 & 0.883305 & 2.4398 & 3.59493 \\
2.100 & 1.19199 & 2.3015 & 3.36535 \\
2.288 & 1.62780 & 1.82527 & 6.10339 \\
2.500 & 1.11638 & 1.36613 & 5.35527 \\
2.857 & 0.30435 & 0.76178 & \\
3.000 &         & 0.54121 & 2.8277 \\
3.102 && 0.45797 & 6.59616 \\
3.106 && 0.45371 & 1.00908 \\
\hline
\end{tabular}
\end{small}

Table 4.1
\end{center}

\newpage
\begin{thebibliography}{99}
\bibitem{ref1} A. Martin, {\it Scattering Theory: Unitarity, Analyticity and
Crossing}, (Springer-Verlag, Berlin, Heidelberg, New York, 1969).
\bibitem{ref2} For a review giving the constraints we use, see S.M.~Roy, Helv.
Physica Acta, {\bf 63} (1990) 627.
\bibitem{ref3} J. Gasser and H. Leutwyler, Ann. Phys. (N.Y.) {\bf 158} (1984)
142.
\bibitem{ref} N. H. Fuchs, H. Sazdjian and J. Stern, Phys. Lett. {\bf B269}
(1991) 183;\\
J. Stern, H. Sazdjian and J. Stern,
Phys. Rev. {\bf D47} (1993) 3814.
\bibitem{ref4} J. Bijnens, G. Colangelo and J. Gasser, University of Bern
preprint BUTP-94/4 (hep-ph/9403390).
\bibitem{ref5} J. Gasser and Ulf-G. Meissner, Phys. Lett. {\bf B258} (1991)
219.
\bibitem{ref6} A. Martin, in {\it High Energy Physics and Elementary
Particles}, p.~155 (International Atomic Energy Agency, Vienna, 1965);
L. Lukaszuk and A. Martin, Nuovo Cimento {\bf 52A} (1967) 122;
C. Lopez and G. Mennessier, Nucl. Phys. {\bf B118} (1977)426.
\bibitem{ref7} Y.S. Jin and A. Martin, Phys. Rev. {\bf 135} (1964) B1369.
\bibitem{ref8} A. Martin, Nuovo Cimento, {\bf 47} (1967) 265; {\bf 58A} (1968)
303.
\bibitem{ref9} G. Auberson, Nuovo Cimento {\bf 68A} (1970) 281;
A.K. Common, Nuovo Cimento {\bf 53A} (1968) 946; {\bf 56A} (1968) 524.
\bibitem{pdg} Particle Data Group, Phys. Rev. {\bf D45} (1992) S1.
\bibitem{mms} B. R. Martin, D. Morgan and G. Shaw,
{\it Pion Pion Interactions in Particle Physics},
(Academic Press, London, New York, San Francisco, 1976).
\bibitem{ref10} G. Auberson, O. Brander, G. Mahoux and A. Martin, Nuovo
Cimento {\bf 65A} (1970) 743.
\bibitem{ref11} P. Grassberger, Nucl. Phys. {\bf B42} (1972) 461.
\bibitem{ref12} A. Martin, Nuovo Cimento, {\bf 63A} (1969) 167.
\end{thebibliography}
\end{document}